%% file: PRL.tex
\documentclass[twocolumn,showpacs,aps,prl,superscriptaddress,floatfix]{revtex4}
\usepackage{epsfig}

\input babarsym
\newcommand{\pmff}[2]{{}^{+#1}_{-#2}{}_{\mathrm{FF}}}
\newcommand{\DBR}{\mbox {$\Delta \ensuremath{\cal B}$}}
\def\BR {\ensuremath{\cal B}\xspace}
\def\ulnu{$b \rightarrow u \ell \nu$}
\def\clnu{$b \rightarrow c \ell \nu$}
\def\xulnu{$B \rightarrow X_u \ell \nu$}

\def\pimunu{$B^{0} \rightarrow \pi^{-} \mu^{+} \nu$}
\def\pilnu{$B^{0} \rightarrow \pi^{-} \ell^{+} \nu$}

\def\rholnu{$B \rightarrow \rho \ell \nu$}
\def\dlnu{$B \rightarrow D \ell \nu$}
\def\dstrlnu{$B \rightarrow D^* \ell \nu$}
\def\bfpilnu{${\BR}(B^{0} \rightarrow \pi^{-} \ell^{+} \nu)$}
\def\bfpilnuq{$\Delta {\BR}(B^{0} \rightarrow \pi^{-} \ell^{+} \nu,q^2)$}

\def\bfpilnuqq{$\DBR(q^2)$}

\def\lnrt{loose neutrino reconstruction technique}

\def\qq{$q^2$}
\def\qqr{$q^2$} 
\def\fplus{$f_+(q^2)$}
\def\vub{$|V_{ub}|$}
\def\VubFpz{$|V_{ub}f_{+}(0)|$}
\def\nQ{12}
\def\udsctau{$u\overline{u}/d\overline{d}/s\overline{s}/c\overline{c}/\tau^+\tau^-$}

\def\obb{other \BB}

\def\DeltaEDef{\ensuremath{\Delta E = (\vec{p}_B \cdot \vec{p}_{beams} - s/2) / \sqrt{s}}\xspace}
\def\FitRegion{\ensuremath{|\Delta E| < 1.0~\gev \mbox{ and } m_{ES} > 5.19~\gev}\xspace}
\newcommand{\gevsq}{\ensuremath{\mathrm{\,Ge\kern -0.1em V^2\!}}\xspace}

\def\sigY{$5072 \pm 251$}
\def\ulnY{$9867 \pm 564$}
\def\obbY{$33341 \pm 409$}
\def\conY{$9299 \pm 450$}   
\def\BFval{$\left(1.46 \pm 0.07_{stat} \pm 0.08_{syst} \right) \times 10^{-4}$}
\def\VubVal{$\left(4.1 \pm 0.2_{stat} \pm 0.2_{syst}\pmff{0.6}{0.4} \right) \times {10^{-3}}$}
\def\VubFpzVal{$\left(9.6 \pm 0.3_{stat} \pm 0.2_{syst} \right) \times {10^{-4}}$}
\def\alphaVal{$0.52 \pm 0.05_{stat} \pm 0.03_{syst}$}
\def\ChiFit{423 for 389 degrees of freedom}
\def\ChiProbBK{65\%}
\def\ChiProbHPQCD{67\%}
\def\ChiProbFNAL{45\%}
\def\ChiProbBZ{41\%}
\def\ChiProbISGW{0.06\%}

\newcommand{\BABARPubYear}{06}
\newcommand{\BABARPubNumber}{069}
\newcommand{\SLACPubNumber}{12253}
\newcommand{\LANLNumber}{0612020}

\begin{document}

\preprint{\babar-PUB-\BABARPubYear/\BABARPubNumber} 
\preprint{SLAC-PUB-\SLACPubNumber} 

\begin{flushleft}
\babar-PUB-\BABARPubYear/\BABARPubNumber\\
SLAC-PUB-\SLACPubNumber\\
hep-ex/\LANLNumber\\
\end{flushleft}

\title{\large\bf\boldmath
Measurement of the \pilnu\ Form-Factor Shape and Branching Fraction,
and Determination of $|V_{ub}|$ with a Loose Neutrino Reconstruction 
Technique}

\input{authors_nov2006}
\date{\today}
\begin{abstract}
 We report the results of a study of the exclusive charmless semileptonic decay,
\pilnu, undertaken with approximately 227 million \BB\ pairs 
collected at the \FourS\ resonance with the \babar\ detector. The analysis uses
events in which the signal $B$ decays are reconstructed with an innovative \lnrt. We 
obtain partial branching fractions in \nQ\ bins of \qq, the momentum transfer
squared, from which we extract the \fplus\ form-factor shape and the total 
branching fraction \bfpilnu\ = \BFval. Based on a recent unquenched lattice QCD calculation 
of the form factor in the range \qq\ $>$ 16 \gevsq, we find the magnitude of 
the CKM matrix element \vub\ to be \VubVal, where the last uncertainty is due to 
the normalization of the form factor.  
\end{abstract}


\maketitle  

 A precise measurement of \vub, the smallest element of the CKM 
matrix~\cite{CKM}, will constrain the description of weak interactions
and \CP\ violation in the Standard Model. The rate for exclusive \pilnu\
decays~\cite{PlusCC} is proportional to $|V_{ub}f_+(q^2)|^2$, where the 
form factor \fplus\ depends on \qq, the momentum transfer squared. Values of
\fplus\ for \pilnu\ decays are provided by unquenched
lattice QCD (LQCD) calculations (HPQCD~\cite{HPQCD04}, FNAL~\cite{FNAL04}), presently reliable only 
at large \qq\ ($>$ 16 \gevsq), by light cone sum rules (LCSR) 
calculations~\cite{ball04}, based on approximations only valid at small 
\qq\ ($<$ 16 \gevsq), and by the ISGW2 quark model calculations~\cite{isgw2}. 
Uncertainties on these calculations dominate the errors on the computed values of \vub. The QCD
theoretical predictions are at present more precise for \pilnu\ than for other 
exclusive \xulnu\ decays, where $X_u$ stands for any charmless meson.
Experimental data can be used to discriminate between 
the various calculations by precisely measuring the \fplus\ shape, thereby 
leading to a smaller theoretical uncertainty on \vub.

 Values of \vub\ have previously been extracted from \pilnu\ measurements by
CLEO~\cite{CLEOpilnu2}, \babar~\cite{PRDJochen, BAD1380} and 
Belle~\cite{BELLEPiRholnu}. In this letter, we present measurements of the
partial branching fractions (BF) \bfpilnuq\ in \nQ\ bins of \qq\ using an innovative
 \lnrt. This leads to more precise values of the total BF \bfpilnu\ and of the 
\fplus\ form-factor shape, which supersede those of our previous untagged measurement~\cite{PRDJochen}. 
We combine the values of \bfpilnuqq\ with recent form-factor 
calculations~\cite{HPQCD04,FNAL04,ball04} to obtain a value of \vub.

 The data set used in this analysis contains approximately 227 million 
\BB\ pairs corresponding to an integrated luminosity of 206~\invfb\ 
collected at the \FourS\ resonance with the \babar\ detector~\cite{ref:babar}
at the \pep2\ asymmetric-energy $e^+e^-$ collider, and of 27.0~\invfb\ integrated 
luminosity of data collected approximately 40 \mev\ below the \FourS\ resonance
(denoted ``off-resonance data''). To estimate the signal efficiency, and the 
signal and background distributions, we use a detailed Monte Carlo (MC) 
simulation of generic \BB\ and \udsctau\ ``continuum'' events as well as \pilnu\
signal events. Signal MC events are produced by the FLATQ2 
generator~\cite{bad809} and are reweighted to reproduce the Becirevic-Kaidalov 
(BK) parametrization~\cite{BK} of $f_+(q^2,\alpha, c_B)$ where 
the values of the shape and normalization parameters, $\alpha$ and $c_B$, are 
taken from Ref.~\cite{PRDJochen}. 
 
 We reconstruct $B$ meson candidates using $\pi^{\pm}$ and 
$\ell^{\mp}$ tracks together with the event's missing momentum $\vec{p}_{miss}$
as an approximation to the signal neutrino momentum. The decay of the second 
$B$ meson is not explicitly reconstructed. The neutrino four-momentum
$P_{miss}\equiv (|\vec{p}_{miss}|,\vec{p}_{miss})$ is inferred from the 
difference between the momentum of the colliding-beam particles $\vec{p}_{beams}$,
and the sum of the momenta of all the charged and neutral particles detected
in a single event $\vec{p}_{tot}$, such that 
$\vec{p}_{miss}\equiv\vec{p}_{beams}-\vec{p}_{tot}$. 
Compared with the tagged analyses in which the two $B$ mesons are
explicitly reconstructed~\cite{BAD1380, BELLEPiRholnu}, the neutrino reconstruction approach
yields a lower signal purity but a significant increase in the signal
reconstruction efficiency. The present \lnrt\ also increases the signal efficiency substantially
with respect to the previous untagged approach by
avoiding the tight neutrino quality cuts~\cite{CLEOpilnu2, PRDJochen} 
which ensure that the neutrino properties are well taken
into account when computing $q^2=(P_{\ell}+P_{\nu})^2$. 
In this analysis, we calculate instead the momentum transfer as
$q^2=(P_{B}-P_{\pi})^2$, where the
ambiguity in the direction of the $B$ meson is handled by use
of the method described in Ref.~\cite{DstrFF}. In this way, the value of \qqr\ is 
unaffected by any mis-reconstruction of the rest of the event.
We obtain a \qq\ resolution of $\sigma$ = 0.52 \gevsq
for the signal candidates in which the pion candidate track truly comes
from a \pilnu\ decay (91\% of the total).
We correct for the reconstruction effects on the \qq\ resolution by applying an 
unregularized unfolding algorithm to the measured \qq\ spectrum~\cite{Cowan}.

 To separate the \pilnu\ signal from the backgrounds, we require two 
well-reconstructed tracks associated with a lepton-pion pair.
The electron (muon) tracks are required to have momenta greater 
than 0.5 (1.0) \gev\ in the laboratory frame to avoid misidentified leptons and 
secondary semileptonic decays. We ensure that the momenta of the lepton and pion 
candidates are kinematically compatible with a real \pilnu\ decay. This requires that a
geometrical vertex fit of the two charged tracks gives a 
$\chi^2$ probability greater than 0.01 and that the angle between the $Y$ and $B$ momenta 
in the \FourS\ frame takes a physical value: $|\cos\theta_{BY}| < 1$, where 
the pseudo-particle $Y$ is defined by its four-momentum $P_Y \equiv (P_{\pi} + P_{\ell})$.
Most backgrounds are efficiently rejected by 
\qqr-dependent cuts on the helicity angle $\theta_{\ell}$ of the $W$ 
boson~\cite{bad809}, on the angle between the thrust axes of the $Y$ and of 
the rest of the event, on the polar angle associated with $\vec{p}_{miss}$, 
and on the squared invariant mass of $P_{miss}$. We reject \pimunu\ candidates with 
$Y$ mass close to the \jpsi\ mass to avoid $\jpsi \rightarrow \mu^+ \mu^-$ decays. 
Non-\BB events are suppressed by requiring the ratio of second to zeroth 
Fox-Wolfram moments to be smaller than 0.5, and by cuts 
\cite{BhabhaVeto} on the number of tracks and clusters. Radiative Bhabha and 
two-photon processes are rejected by vetoing events containing a photon 
conversion and by requiring $(\vec{p}_{tot}\cdot\hat{z})/E_{tot} < 0.64$ 
and $(\vec{p}_{tot}\cdot\hat{z})/E_{tot} > 0.35$ for candidates in the 
electron and positron channels, respectively, where the $z$ axis is given by 
the electron beam direction. We reduce the remaining backgrounds with the 
variables \DeltaEDef and $\ensuremath{m_{ES} = \sqrt{(s/2+\vec{p}_B \cdot 
\vec{p}_{beams})^2/E_{beams}^2 - \vec{p}_B^{\,2}}}$, where 
$\vec{p}_B = \vec{p}_{\pi} + \vec{p}_{\ell} + \vec{p}_{miss}$ and
$\sqrt{s}$ is the total energy in the \FourS\ frame. Only 
candidates with \FitRegion\ are retained. When several candidates remain in an 
event after these cuts, the candidate with $\cos\theta_{\ell}$ closest to 
zero is selected. This rejects 30\% of the combinatorial 
signal candidates while keeping 97\% of the correct ones. The signal 
event reconstruction efficiency varies between 6.7\% and
9.8\%, depending on the \qq\ bin. 

\begin{figure}
\begin{center}
\epsfig{file=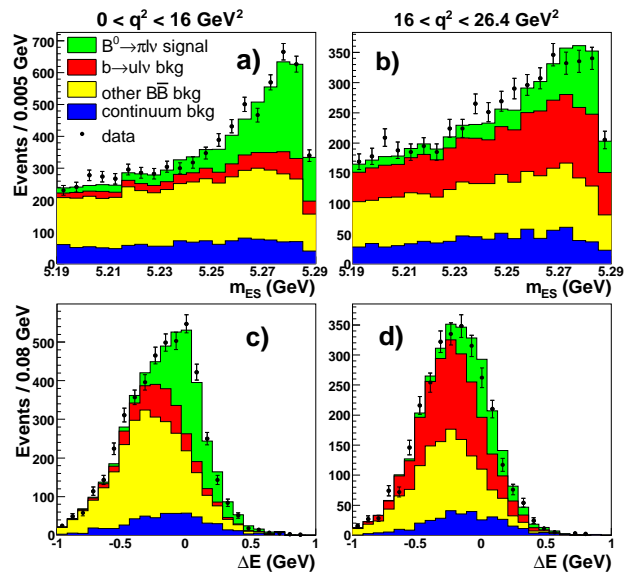,height=7.6cm}
\caption[]{\label{dEmESProj} Yield fit projections for (a,b) \mes\ with  
$-0.16 < \Delta E < 0.20$ \gev; and (c,d) \DeltaE\ with \mes\ $>$ 5.272 \gev.
The distributions (a,c) are for \qqr\ $<$ 16 \gevsq; and (b,d) are for \qqr\ $>$ 16 \gevsq.}
\end{center}
\end{figure}

The \pilnu\ signal yield is obtained as a function of \qqr\ by
performing a two-dimensional extended maximum-likelihood fit~\cite{Barlow} on 
\mes, and \DeltaE\ in each bin of \qqr. The data samples in each \qqr\ bin are divided into 
four categories: \pilnu\ signal, other \ulnu, \obb, and continuum backgrounds. 
These four types of events have distinct structures in the two-dimensional 
\mes--\DeltaE\ plane. We use the \mes--\DeltaE\ histograms obtained from the MC 
simulation as two-dimensional probability density functions (PDFs). 
The yields of the signal, \ulnu\ background and \obb\ background,
subdivided in twelve, three and four \qqr\ bins, respectively,
are extracted from a nineteen-parameter fit of the MC PDFs to the experimental data.
The continuum background is corrected to
match the off-resonance data control sample and is fixed in the fit. The number 
and type of fit parameters were chosen to provide a good balance between 
reliance on simulation predictions, complexity of the fit and total error size.
\mes\ and \DeltaE\ fit projections for the experimental data are shown in 
Fig.~\ref{dEmESProj} in two ranges of \qq\ corresponding to the sum of eight bins below and
four bins above \qqr\ = 16 \gevsq. We obtain \sigY\ 
events for the total signal yield, \ulnY\ events for the \ulnu\ background, 
\obbY\ events for the \obb\ backgrounds, and \conY\ events for the continuum 
yield. The fit has a $\chi^2$ value of \ChiFit.

Numerous sources of systematic uncertainties and their correlations among the \qq\ bins have been 
investigated. The uncertainties due to the detector simulation are established 
by varying within bounds given by control samples the tracking efficiency of all 
charged tracks, the particle identification efficiencies of signal candidate 
tracks, the calorimeter efficiency (varied separately for photons, \KL and 
neutrons) and the energy deposited in the calorimeter by \KL mesons. 
The reconstruction of these neutral particles affects the analysis via
the neutrino reconstruction. The 
uncertainties due to the generator-level inputs to the simulation are established
by varying, within errors~\cite{PDG06}, the BFs of the background processes 
\ulnu, \clnu, $D \rightarrow X \ell \nu$ and $D \rightarrow K^0_L X$ as well as 
the BF of the $\Upsilon(4S) \rightarrow \BzBzb$ decay. The \pilnu, \rholnu,
\dlnu\ and \dstrlnu\ form factors are varied within bounds given by recent 
calculations~\cite{ball05} or measurements~\cite{DstrFF,PrelimRes,PDG06}. The heavy quark 
parameters used in the simulation of non-resonant \ulnu\ events are varied 
according to Ref.~\cite{Henning}. We assign an uncertainty of 
20\% to the final state radiation (FSR) corrections calculated by PHOTOS~\cite{photos, photosErr}. 
Finally, the uncertainties due to the modeling of the continuum are established 
by varying its \qqr, \mes, and \DeltaE\ shapes and total yield 
within their errors given by comparisons with the off-resonance data control sample. 
The high statistics provided by our technique allow us to show that there is good 
agreement between data and simulation for the critical variables in signal depleted, 
signal enhanced, \ulnu\ enhanced and continuum control samples. 
Consistent results are obtained either by dividing the final dataset into
sub-samples or using modified binnings or modified event selections.

 The partial BFs are calculated using the observed signal yields, the unfolding 
algorithm and the signal efficiencies given by the simulation. The total BF is 
given by the sum of the partial BFs, thereby reducing the sensitivity of 
the signal efficiency to the uncertainties of the \fplus\ form factor. We compute
the covariance matrix for each source of uncertainty and use these matrices to 
calculate the errors on the total BF. The fit and systematic errors are given 
in Table \ref{errors} for five ranges of \qq. The complete set of fit and 
systematic uncertainties of the partial and total BFs as well as their 
correlation matrices are given in Ref.~\cite{EPAPS}. Our value of the total
BF, \BFval, is comparable in precision to the world average prior to our 
result~\cite{PDG06}: $(1.35\pm0.08_{stat}\pm0.08_{syst}) \times 10^{-4}$.
The systematic error is due in large part to the detector efficiency. 
The systematic errors arising from the BFs and form factors of the backgrounds 
have been reduced with respect to previous untagged measurements by 
the many-parameter fit to the background yields in the 12 bins of \qqr. 

\begin{table}
\begin{center}
\caption[]{\label{errors} Values of \bfpilnuqq\ and their relative errors (\%).}
\begin{tabular}{lccccc} 
\hline\hline \qq\ bins (\gevsq) & 4--6 & 16--18 &\qq$<$16 &\qq$>$16 & full \qq\ range \\ 
\hline 
BF ($10^{-4}$)                   & 0.16 & 0.13 & 1.09 & 0.38 & 1.46 \\
\hline
Fit error                        & 12.8 & 17.6 & 5.3 & 10.3 & 4.8 \\ 
Detector effects                 & 3.7 & 5.0 & 4.4 & 4.5 & 3.7 \\ 
Continuum bkg                    & 1.2 & 1.7 & 2.8 & 3.5 & 2.5 \\ 
$B \rightarrow X_u \ell \nu$ bkg & 3.0 & 3.1 & 2.3 & 4.7 & 2.5 \\ 
$B \rightarrow X_c \ell \nu$ bkg & 1.7 & 1.8 & 1.2 & 1.2 & 1.0 \\ 
Other effects                    & 3.4 & 3.0 & 2.3 & 2.3 & 2.3 \\ 
\hline
Total error                      & 14.2 & 19.0 & 8.2 & 12.9 & 7.5 \\ 

\hline\hline 
\end{tabular}
\end{center}
\end{table}

\begin{figure}[!htb]
\begin{center}
\epsfig{file=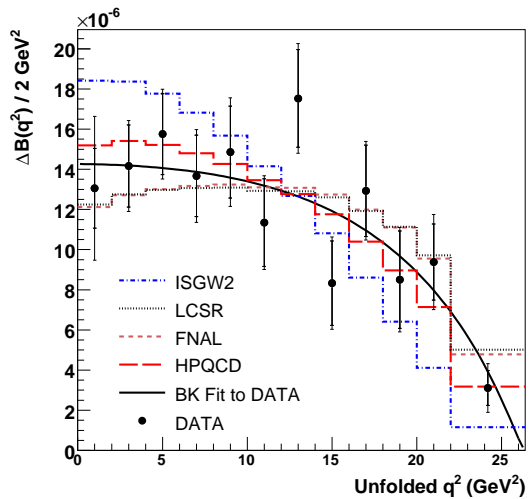,height=7cm}
\caption[]{\label{figFplus} 
Partial \bfpilnuqq\ spectrum in 12 bins of \qq. The smaller error 
bars are statistical only while the larger ones also include systematic 
uncertainties. The solid black curve shows the result of the fit of the BK parametrization to the data.
The data are also compared to unquenched LQCD 
calculations~\cite{HPQCD04,FNAL04}, LCSR calculations~\cite{ball04}, and the 
ISGW2 quark model~\cite{isgw2}.}  
\end{center}
\end{figure}

The \bfpilnuqq\ distribution is displayed in Fig.~\ref{figFplus} together with 
theoretical predictions. 
We modify the measured \qq\ distribution to remove FSR effects, in order 
to allow a direct comparison with the theoretical predictions 
which do not include such effects (this procedure is referred to as ``No FSR'' in Ref.~\cite{EPAPS}).
We obtain the \fplus\ shape from a fit to this distribution. The 
$\chi^2$ function minimized in the \fplus\ fit uses a PDF based on the two-parameter
BK parametrization. It is defined in terms of the \bfpilnuqq\ covariance matrix 
to take into account the correlations among the measurements in the various 
\qq\ bins. The fit gives $\alpha$ = \alphaVal, compared to our previous untagged 
measurement $\alpha = 0.61 \pm 0.09$~\cite{PRDJochen} (statistical error only)
as well as a value of \VubFpz\ = \VubFpzVal\ from the fit extrapolated to \qq\ $= 0$,
with $P(\chi^2)$=\ChiProbBK.
This value includes a 67\% anti-correlation between the shape and normalization parameters,
$\alpha$ and $c_B$, and can be used to predict~\cite{Hill} rates of other decays such as 
$B \rightarrow \pi \pi$. 

The $\chi^2$ probabilities have been calculated relative to the binned 
data result for various theoretical predictions, considering only 
experimental errors. We obtain $P(\chi^2)$=\ChiProbHPQCD\ for 
HPQCD~\cite{HPQCD04}, \ChiProbFNAL\ for FNAL~\cite{FNAL04} and 
\ChiProbBZ\ for LCSR~\cite{ball04}. The ISGW2 quark model~\cite{isgw2}, 
$P(\chi^2)$=\ChiProbISGW, is clearly incompatible with our data.

\begin{table}
\caption[]{\label{vubtable}Values of $|V_{ub}|$ derived 
from form-factor calculations. The first two errors arise from the 
statistical and systematic uncertainties of the partial BFs, respectively. 
The third error comes from the uncertainty on $\Delta \zeta$.} 
\begin{center}
\begin{tabular}{lccc}
\hline\hline
                  & $q^2$ (\gevsq) & $\Delta\zeta$ (ps$^{-1}$) & $|V_{ub}|$ ($10^{-3}$)\\ \hline
HPQCD~\cite{HPQCD04}      & $>$ 16 & $1.46\pm0.35$ & $4.1\pm0.2\pm0.2~{}^{+0.6}_{-0.4}$\\
FNAL~\cite{FNAL04}        & $>$ 16 & $1.83\pm0.50$ & $3.7\pm0.2\pm0.2~{}^{+0.6}_{-0.4}$\\
LCSR~\cite{ball04}        & $<$ 16 & $5.44\pm1.43$ & $3.6\pm0.1\pm0.1~{}^{+0.6}_{-0.4}$\\
ISGW2~\cite{isgw2}        & 0--26.4 & $9.6\pm4.8$   & $3.2\pm0.1\pm0.1~{}^{+1.3}_{-0.6}$\\
\hline\hline
\end{tabular}
\end{center}
\end{table}

We extract \vub\ from the partial BFs \bfpilnuqq\ using the relation:
$|V_{ub}| = \sqrt{\Delta{\cal B}(q^2)/(\tau_{B^0}\Delta \zeta)}$, where
$\tau_{B^0} = 1.530 \pm 0.009$ ps~\cite{PDG06} is the $B^0$ lifetime and 
$\Delta\zeta = \Gamma/|V_{ub}|^2$ is the normalized partial decay rate
predicted by the form-factor calculations~\cite{HPQCD04, FNAL04, ball04, isgw2}. 
Excluding the ISGW2 model, the values of \vub\ given in
Table \ref{vubtable}\ range from $(3.6-4.1)\times 10^{-3}$. 
 
In summary, we have measured the partial \pilnu\ branching fractions in \nQ\ 
bins of \qq\ using a \lnrt. We obtained the most precise measurement to 
date of the \bfpilnu\ and \VubFpz, as well as a detailed description of the \fplus\ shape. 
This shape can be compared with various theoretical predictions and, in 
particular, shows that the ISGW2 model can be ruled out. From the most 
recently published unquenched LQCD calculation~\cite{HPQCD04}, we obtain 
\vub\ = \VubVal.

\input{acknow_PRL}

\begin{table*}
\flushleft
\textbf{\large Electronic Physics Auxiliary Publication Service (EPAPS)}\\
\smallskip
\normalsize
This is an EPAPS attachment to
B. Aubert \textit{et al.} (\babar\ Collaboration),
\babar-PUB-\BABARPubYear/\BABARPubNumber,
SLAC-PUB-\SLACPubNumber,
hep-ex/\LANLNumber,
submitted to Phys.\ Rev.\ Lett.
For more information on EPAPS, see
http://www.aip.org/pubservs/epaps.html.
\end{table*}

\renewcommand{\thetable}{A-\arabic{table}}
\setcounter{table}{0}

\begin{table*}
\caption[]{\label{TabMainMeas} \pilnu\ yields, efficiency (\%), \DBR\ ($10^{-7}$) and their relative errors (\%). The \DBR\ and 
efficiency values labelled ``No FSR'' are modified to remove FSR effects.
This procedure has no significant impact on the \DBR\ values.}
\begin{small}
\begin{center}
\begin{tabular}{lccccccccccccccc}
\hline\hline
\qq\ bins (\gevsq) & 0-2 & 2-4 & 4-6 & 6-8 & 8-10 & 10-12 & 12-14 &14-16&16-18&18-20&20-22&22-26.4&\qq$<$16&\qq$>$16&Total\\ \hline
Fitted yield          & 366.6& 462.9& 499.5& 451.8& 436.4& 391.0& 522.7& 333.6& 458.0& 355.4& 364.8& 428.8& 3464.6& 1606.9& 5071.5 \\ 
Fit error & 12.9& 9.7& 8.6& 9.7& 11.2& 13.0& 11.7& 17.4& 15.6& 21.7& 15.9& 17.3& 5.3& 9.9& 5.0 \\ 
Systematic error  & 20.0& 6.3& 3.2& 4.9& 6.3& 4.1& 4.3& 6.1& 4.9& 7.9& 12.3& 17.4& 3.7& 7.3& 3.9 \\ 
\hline
Unfolded yield       & 374.7& 452.3& 515.3& 442.2& 459.1& 360.7& 583.4& 302.7& 514.3& 357.7& 406.3& 303.0& 3490.2& 1581.3& 5071.5  \\ 
Fit error & 15.2& 14.4& 12.8& 14.8& 15.4& 19.2& 13.9& 25.2& 17.6& 28.5& 20.2& 27.7& 5.4& 10.2& 5.0 \\ 
Systematic error    & 22.9& 7.3& 3.7& 5.6& 7.8& 5.2& 4.8& 8.9& 5.2& 10.0& 14.9& 27.3& 3.7& 7.6& 3.9 \\ 
\hline
Efficiency  & 6.56 & 7.13 & 7.22 & 7.11 & 6.76 & 6.97 & 7.21 & 7.87 & 8.68 & 9.20 & 9.37 & 9.66 &-&-&- \\
Eff. (No FSR)    & 6.31 & 7.02 & 7.19 & 7.11 & 6.79 & 6.99 & 7.32 & 7.99 & 8.75 & 9.25 & 9.53 & 9.73 &-&-&- \\ 
\hline
\DBR\  & 125.5& 139.5& 156.9& 136.8& 149.4& 113.7& 177.9& 84.5& 130.3& 85.5& 95.3& 68.9& 1084.3& 380.0& 1464.3   \\ 
\DBR\ (No FSR) & 130.6& 141.6& 157.5& 136.7& 148.6& 113.5& 175.3& 83.3& 129.3& 85.1& 93.8& 68.4& 1087.1& 376.6& 1463.7 \\
Fit error & 15.2& 14.4& 12.8& 14.8& 15.4& 19.2& 13.9& 25.2& 17.6& 28.5& 20.2& 27.7& 5.3& 10.3& 4.8 \\ 
Systematic error  & 23.7& 7.0& 6.2& 8.1& 9.6& 7.3& 7.1& 11.0& 7.0& 11.0& 14.9& 27.0& 6.3& 7.8& 5.7 \\ 
\hline\hline
\end{tabular}
\end{center}
\end{small}
\end{table*}

\begin{table*}
\caption[]{\label{BFErrTable} Relative errors (\%) of the partial and total \bfpilnu\ from all sources. FSR effects are included.}
\begin{small}
\begin{center}
\begin{tabular}{lccccccccccccccc}
\hline\hline
\qq\ bins (\gevsq) & 0-2 & 2-4 & 4-6 & 6-8 & 8-10 & 10-12 & 12-14 & 14-16 & 16-18& 18-20& 20-22& 22-26.4& \qq$<$16 & \qq$>$16 & Total \\ 
\hline
Tracking efficiency   & 1.6& 1.7& 1.3& 3.1& 3.8& 1.3& 1.8& 7.1& 2.3& 1.7& 2.2& 9.2& 1.9& 1.8& 1.1 \\ 
$\gamma$ efficiency   & 4.7& 1.3& 2.6& 5.0& 3.6& 3.2& 3.5& 3.1& 3.0& 3.5& 3.8& 7.0& 2.9& 1.7& 1.9 \\ 
\KL\ \& neutrons      & 0.7& 0.6& 0.7& 1.0& 1.2& 1.2& 1.2& 1.3& 2.0& 1.5& 1.1& 2.2& 0.5& 1.0& 0.6 \\ 
Particle ID           & 7.0& 2.5& 2.1& 1.9& 0.7& 2.7& 2.6& 2.5& 2.6& 3.4& 2.9& 7.0& 2.6& 3.6& 2.9 \\ 
\hline
Continuum yield       & 7.0& 0.5& 0.6& 0.2& 0.9& 1.0& 0.6& 0.8& 0.6& 1.9& 1.1& 4.0& 1.0& 1.6& 1.0 \\ 
Continuum \qqr        & 20.1& 1.5& 1.0& 1.0& 1.7& 1.8& 1.5& 2.0& 1.5& 3.3& 4.0& 8.7& 2.4& 1.9& 1.8 \\ 
Continuum \mes        & 1.1& 0.3& 0.1& 0.1& 0.1& 0.1& 0.3& 0.6& 0.3& 0.6& 0.7& 1.0& 0.2& 0.5& 0.2 \\ 
Continuum \DeltaE     & 3.0& 1.6& 0.5& 1.0& 1.5& 0.2& 0.1& 1.2& 0.6& 1.8& 3.8& 5.2& 1.0& 2.5& 1.4 \\ 
\hline
$B\rightarrow X_u\ell\nu$ BFs & 1.2& 1.4& 0.7& 0.7& 1.0& 1.4& 1.1& 1.8& 1.7& 3.6& 10.4& 12.1& 0.9& 3.4& 1.2 \\ 
SF param              & 0.4& 0.5& 0.3& 0.5& 0.1& 0.1& 0.1& 0.4& 0.4& 4.1& 5.7& 14.9& 0.2& 2.1& 0.7 \\ 
\rholnu\ FFs          & 1.3& 0.7& 1.8& 1.1& 0.8& 1.1& 1.2& 3.2& 0.5& 3.3& 1.3& 4.3& 0.9& 0.8& 0.6 \\ 
\pilnu\ FF            & 0.4& 0.5& 0.5& 0.5& 0.5& 0.5& 0.5& 0.8& 0.7& 0.7& 1.1& 3.5& 0.5& 1.3& 0.7 \\ 
FSR                   & 0.7& 1.5& 2.2& 1.9& 2.6& 2.7& 2.2& 0.6& 2.5& 2.0& 1.3& 1.1& 1.9& 1.8& 1.9 \\ 
\hline
$B\rightarrow X_c\ell\nu$ BFs & 1.8& 2.1& 1.1& 2.2& 4.6& 1.2& 2.4& 2.2& 1.3& 2.5& 1.7& 2.2& 0.9& 1.0& 0.8 \\ 
\dstrlnu\ FFs         & 0.7& 1.1& 0.1& 1.6& 3.1& 0.8& 1.4& 1.0& 0.9& 1.2& 0.6& 2.7& 0.7& 0.4& 0.6 \\ 
\dlnu\ FF             & 1.7& 1.2& 0.7& 0.3& 2.2& 0.4& 0.1& 0.7& 0.6& 0.9& 0.4& 0.7& 0.1& 0.4& 0.2 \\ 
\hline
$\Upsilon(4S)\rightarrow \BzBzb$ BF & 2.1& 2.5& 1.5& 1.5& 1.3& 1.7& 1.6& 1.2& 1.6& 1.0& 2.4& 1.9& 1.7& 1.7& 1.7 \\ 
$D\rightarrow X\ell\nu$ BFs & 2.3& 2.8& 1.1& 1.3& 1.6& 1.1& 1.1& 0.9& 0.6& 0.9& 1.0& 0.8& 0.4& 0.5& 0.3 \\ 
$D\rightarrow K^0_L$ BFs & 0.6& 1.7& 2.3& 1.5& 2.0& 2.3& 1.0& 4.2& 1.8& 3.9& 1.0& 1.7& 1.1& 1.0& 1.1 \\ 
B counting            & 1.1& 1.1& 1.1& 1.1& 1.1& 1.1& 1.1& 1.1& 1.1& 1.1& 1.1& 1.1& 1.1& 1.1& 1.1 \\ 
Signal MC stat error  & 1.5& 1.7& 1.6& 1.9& 1.6& 1.9& 1.4& 1.8& 1.4& 1.6& 1.3& 1.5& 0.5& 0.6& 0.4 \\ 
\hline\hline
Total systematic error          & 23.7& 7.0& 6.2& 8.1& 9.6& 7.3& 7.1& 11.0& 7.0& 11.0& 14.9& 27.0& 6.3& 7.8& 5.7 \\ 
Fit error & 15.2& 14.4& 12.8& 14.8& 15.4& 19.2& 13.9& 25.2& 17.6& 28.5& 20.2& 27.7& 5.3& 10.3& 4.8 \\ 
Total error & 28.2& 16.1& 14.2& 16.9& 18.2& 20.5& 15.6& 27.5& 19.0& 30.6& 25.1& 38.7& 8.2& 12.9& 7.5\\ 
\hline\hline
\end{tabular}
\end{center}
\end{small}
\end{table*}

\begin{table*}
\caption[]{\label{StatCovMC}Correlation matrix of the partial \bfpilnuq\ statistical errors. 
The correlations have the same values for the ``No FSR'' case as for the one with FSR, within the quoted precision.}
\begin{center}
\begin{small}
\begin{tabular}{r|cccccccccccc}
\qq\ bins \\
(\gevsq) & 0-2 & 2-4 & 4-6 & 6-8 & 8-10 & 10-12 & 12-14 & 14-16 & 16-18& 18-20 & 20-22 & 22-26.4\\ \hline
0-2 & 1.00 & -0.26 & 0.11 & 0.01 & 0.06 & 0.01 & 0.03 & -0.01 & -0.00 & -0.00 & -0.00 & -0.01\\
2-4 & -0.26 & 1.00 & -0.33 & 0.14 & 0.03 & -0.00 & 0.01 & -0.00 & -0.00 & -0.00 & -0.00 & -0.00\\
4-6 & 0.11 & -0.33 & 1.00 & -0.30 & 0.21 & 0.05 & 0.13 & -0.02 & -0.00 & -0.00 & -0.00 & -0.00\\
6-8 & 0.01 & 0.14 & -0.30 & 1.00 & -0.22 & 0.15 & 0.09 & -0.01 & -0.00 & -0.00 & -0.00 & -0.00\\
8-10 & 0.06 & 0.03 & 0.21 & -0.22 & 1.00 & -0.22 & 0.20 & -0.03 & 0.00 & -0.00 & -0.00 & -0.00\\
10-12 & 0.01 & -0.00 & 0.05 & 0.15 & -0.22 & 1.00 & -0.02 & 0.02 & -0.00 & 0.00 & -0.00 & -0.00\\
12-14 & 0.03 & 0.01 & 0.13 & 0.09 & 0.20 & -0.02 & 1.00 & -0.25 & -0.00 & -0.03 & 0.00 & -0.00\\
14-16 & -0.01 & -0.00 & -0.02 & -0.01 & -0.03 & 0.02 & -0.25 & 1.00 & 0.06 & 0.21 & -0.06 & -0.04\\
16-18 & -0.00 & -0.00 & -0.00 & -0.00 & 0.00 & -0.00 & -0.00 & 0.06 & 1.00 & 0.13 & -0.08 & -0.06\\
18-20 & -0.00 & -0.00 & -0.00 & -0.00 & -0.00 & 0.00 & -0.03 & 0.21 & 0.13 & 1.00 & -0.21 & -0.13\\
20-22 & -0.00 & -0.00 & -0.00 & -0.00 & -0.00 & -0.00 & 0.00 & -0.06 & -0.08 & -0.21 & 1.00 & -0.05\\
22-26.4 & -0.01 & -0.00 & -0.00 & -0.00 & -0.00 & -0.00 & -0.00 & -0.04 & -0.06 & -0.13 & -0.05 & 1.00\\
\end{tabular}
\end{small}
\end{center}
\end{table*}

\begin{table*}
\caption[]{\label{SystCovMC}Correlation matrix of the partial \bfpilnuq\ systematic errors. 
The correlations have the same values for the ``No FSR'' case as for the one with FSR, within the quoted precision.}
\begin{center}
\begin{small}
\begin{tabular}{r|cccccccccccc}
\qq\ bins \\
(\gevsq) & 0-2 & 2-4 & 4-6 & 6-8 & 8-10 & 10-12 & 12-14 & 14-16 & 16-18& 18-20 & 20-22 & 22-26.4\\ \hline
0-2 & 1.00 & 0.19 & 0.32 & 0.11 & -0.06 & 0.46 & 0.44 & 0.13 & 0.31 & 0.23 & 0.13 & 0.00\\
2-4 & 0.19 & 1.00 & 0.21 & -0.09 & -0.28 & 0.31 & 0.11 & -0.05 & 0.23 & 0.14 & 0.18 & 0.35\\
4-6 & 0.32 & 0.21 & 1.00 & 0.66 & 0.46 & 0.74 & 0.58 & 0.52 & 0.56 & 0.30 & 0.04 & 0.04\\
6-8 & 0.11 & -0.09 & 0.66 & 1.00 & 0.75 & 0.58 & 0.67 & 0.60 & 0.54 & 0.27 & -0.05 & -0.09\\
8-10 & -0.06 & -0.28 & 0.46 & 0.75 & 1.00 & 0.32 & 0.59 & 0.48 & 0.35 & 0.13 & 0.04 & -0.11\\
10-12 & 0.46 & 0.31 & 0.74 & 0.58 & 0.32 & 1.00 & 0.67 & 0.37 & 0.55 & 0.36 & 0.08 & 0.05\\
12-14 & 0.44 & 0.11 & 0.58 & 0.67 & 0.59 & 0.67 & 1.00 & 0.32 & 0.62 & 0.36 & 0.08 & -0.14\\
14-16 & 0.13 & -0.05 & 0.52 & 0.60 & 0.48 & 0.37 & 0.32 & 1.00 & 0.40 & 0.28 & 0.05 & -0.11\\
16-18 & 0.31 & 0.23 & 0.56 & 0.54 & 0.35 & 0.55 & 0.62 & 0.40 & 1.00 & 0.54 & 0.05 & -0.08\\
18-20 & 0.23 & 0.14 & 0.30 & 0.27 & 0.13 & 0.36 & 0.36 & 0.28 & 0.54 & 1.00 & -0.10 & 0.23\\
20-22 & 0.13 & 0.18 & 0.04 & -0.05 & 0.04 & 0.08 & 0.08 & 0.05 & 0.05 & -0.10 & 1.00 & 0.08\\
22-26.4 & 0.00 & 0.35 & 0.04 & -0.09 & -0.11 & 0.05 & -0.14 & -0.11 & -0.08 & 0.23 & 0.08 & 1.00\\
\end{tabular}
\end{small}
\end{center}
\end{table*}

\end{document}

%% file: authors_nov2006.tex
%
\author{B.~Aubert}
\author{M.~Bona}
\author{D.~Boutigny}
\author{Y.~Karyotakis}
\author{J.~P.~Lees}
\author{V.~Poireau}
\author{X.~Prudent}
\author{V.~Tisserand}
\author{A.~Zghiche}
\affiliation{Laboratoire de Physique des Particules, IN2P3/CNRS et Universit\'e de Savoie, F-74941 Annecy-Le-Vieux, France }
\author{E.~Grauges}
\affiliation{Universitat de Barcelona, Facultat de Fisica, Departament ECM, E-08028 Barcelona, Spain }
% ...
\collaboration{The \babar\ Collaboration}
\noaffiliation

%% file: acknow_PRL.tex
We are grateful for the excellent luminosity and machine conditions
provided by our \pep2\ colleagues, 
and for the substantial dedicated effort from
the computing organizations that support \babar.
The collaborating institutions wish to thank 
SLAC for its support and kind hospitality. 
This work is supported by
DOE
and NSF (USA),
NSERC (Canada),
IHEP (China),
CEA and
CNRS-IN2P3
(France),
BMBF and DFG
(Germany),
INFN (Italy),
FOM (The Netherlands),
NFR (Norway),
MIST (Russia),
MEC (Spain), and
PPARC (United Kingdom). 
Individuals have received support from the
Marie Curie EIF (European Union) and
the A.~P.~Sloan Foundation.

%% file: PRL.bbl
\begin{thebibliography}{99}

\bibitem{CKM}
M. Kobayashi and T. Maskawa, Prog. Theor. Phys. {\bf{49}}, 652 (1973).

\bibitem{PlusCC}
Charge conjugate decays and $\ell=e$ or $\mu$ are implied throughout this
paper.

\bibitem{HPQCD04}
E. Gulez {\em et al.} (HPQCD Collaboration), Phys. Rev. {\bf D73}, 074502 (2006).

\bibitem{FNAL04}
M. Okamoto {\em et al.}, Nucl. Phys. Proc. Suppl. {\bf 140}, 461 (2005).

\bibitem{ball04}
P. Ball and R. Zwicky, Phys. Rev. {\bf D71}, 014015 (2005).

\bibitem{isgw2}
D. Scora and N. Isgur, Phys. Rev. {\bf D52}, 2783 (1995).

\bibitem{CLEOpilnu2}
S. B. Athar {\it et al.} (CLEO Collaboration), Phys. Rev. {\bf D68}, 072003 (2003).

\bibitem{PRDJochen}
B. Aubert {\it et al.} (\babar\ Collaboration), Phys. Rev. {\bf D72}, 051102 (2005).

\bibitem{BAD1380}
B. Aubert {\it et al.} (\babar\ Collaboration), Phys. Rev. Lett. {\bf 97}, 211801 (2006).

\bibitem{BELLEPiRholnu}
T. Hokuue {\em et al.} (Belle Collaboration), hep-ex/0604024, submitted to Phys. Lett. B.

\bibitem{ref:babar}
B.\ Aubert {\em et al.} (\babar\ Collaboration), Nucl.\ Instrum.\ Methods {\bf A479}, 1 (2002).

\bibitem{bad809}
D. C\^ot\'e {\em et al.}, Eur.\ Phys.\ J.\ C {\bf 38}, 105 (2004).

\bibitem{BK}
D. Becirevic and A. B. Kaidalov, Phys. Lett. {\bf B478}, 417 (2000).

\bibitem{DstrFF}
B. Aubert {\it et al.} (\babar\ Collaboration), Phys. Rev. {\bf D74}, 092004 (2006).

\bibitem{Cowan}
G. Cowan, Statistical Data Analysis, Chap. 11, Oxford University Press (1998).

\bibitem{BhabhaVeto} 
B. Aubert {\it et al.} (\babar\ Collaboration), Phys. Rev. {\bf D67}, 031101 (2003).

\bibitem{Barlow}
R.J. Barlow and C. Beeston, Comput. Phys. Commun. \textbf{77}, 219 (1993).

\bibitem{PDG06}
W.-M. Yao {\em et al.} (Particle Data Group), \jpg{33}, 1 (2006).

\bibitem{ball05}
P. Ball and R. Zwicky, Phys. Rev. {\bf D71}, 014029 (2005).

\bibitem{PrelimRes}
B. Aubert {\it et al.} (\babar\ Collaboration), hep-ex/0607060.

\bibitem{Henning}
O. L. Buchm\"uller and H. U. Fl\"acher, Phys. Rev. {\bf D73}, 073008 (2006).

\bibitem{photos}
E. Barberio and Z. Was, Comput. Phys. Commun. {\bf 79}, 291 (1994).

\bibitem{photosErr}
E. Richter-Was {\em et al.}, Phys. Lett. {\bf B303}, 163 (1993).

\bibitem{EPAPS}
See EPAPS Document No.\ $xxxxxx$ for tables of systematic errors
and error matrices. For more information on EPAPS, see http://www.aip.org/pubservs/epaps.html. 

\bibitem{Hill}
T. Becher and R. J. Hill, Phys. Lett. {\bf B633}, 61 (2006).

\end{thebibliography}
